\documentclass[pre,twocolumn, superscriptaddress]{revtex4-1}
\usepackage{amsmath}
\usepackage{amsfonts}
\usepackage{amssymb}
\usepackage{array}
\usepackage{dcolumn}
\usepackage{longtable}
\usepackage{hyperref}

\usepackage{float}
\usepackage{bm}
\usepackage{graphicx}
\usepackage{natbib}
\usepackage{color}

\newcommand{\Up}{\mathbf{U}}
\newcommand{\Dn}{\mathbf{D}}

\newcommand{\ie}{\textit{i.e.}\ }
\newcommand{\eg}{\textit{e.g.}\ }
\newcommand{\viz}{\textit{viz.}\ }

\newcommand{\muchlessthan}{\ll}
\newcommand{\definedas}{\equiv}
\newcommand{\goesas}{\sim}

\begin{document}
\title{Cyclic annealing as an iterated random map}
\date{\today}
\author{Muhittin Mungan}

\email{mungan@iam.uni-bonn.de}
\affiliation{Institut f\"{u}r angewandte Mathematik, Universit\"{a}t Bonn, Endenicher Allee 60, 53115 Bonn, Germany}
\author{Thomas A. Witten}
\email{t-witten@uchicago.edu}
\affiliation{Department of Physics and James Franck Institute, University of Chicago, Chicago, Illinois 60637, United States.}

\begin{abstract}

    Disordered magnets, martensitic mixed crystals, and glassy solids can be irreversibly deformed by subjecting them to external deformation.  The deformation produces a smooth, reversible response punctuated by abrupt relaxation ``glitches".  Under appropriate repeated forward and reverse deformation producing multiple glitches, a strict repetition of a single sequence of microscopic configurations often emerges.  We exhibit these features by describing the evolution of the system configuration from glitch to glitch as a mapping of $\mathcal{N}$ states into one-another.  A map $\Up$ controls forward deformation; a second map $\Dn$ controls reverse deformation.  Iteration of a given sequence of forward and reverse maps, \eg $\Dn\Dn\Dn\Dn\Up\Up\Up\Up$ necessarily produces a convergence to a fixed cyclic repetition of states covering multiple glitches. The repetition may have a period of more than one strain cycle, as recently observed in simulations.  
    Using numerical sampling, we characterize the convergence properties of four types of random maps implementing successive physical restrictions. The most restrictive is the much-studied Preisach model. These maps show only the most qualitative resemblance to annealing simulations. However, they suggest further properties 
    needed for a realistic mapping scheme. 
    
\end{abstract}
\maketitle

    \section{Introduction}\label{sec:introduction}
    Much recent interest has focused on cyclic annealing of disordered materials, \ie deforming them repeatedly in a prescribed way.  Example materials are 
    magnets, dense colloidal suspensions, sheared amorphous solids or granular packs, for which experimental and numerical results have been reported \cite{Sethna93,Chaikin2008,Keimetal2011,Keimetal2013,Keimetal2014,Regev2013,Regev2015, Fioccoetal2014, Fiocco2015,Sastry2017,adhikari2018memory,KeimArratia2015,mukherji2018strength,Royer49,keim2018return, keim2018memory}. As these materials are deformed by increasing amounts, they suffer a series of abrupt yielding events that move them discontinuously to new internal states.  Repeated large-amplitude cycles of deformation cause many such yielding events in each cycle; these events differ with every cycle as new internal configurations are encountered.  However, it often happens that this annealing process reaches an end point at which every subsequent cycle leads to the same sequence of yielding events involving the same sequence of configurations.  Thus the deformation process has led the system to a small family of specific states from among the much larger number of available ones.  Our aim in this paper is to explore a minimal mechanism for the attainment of this order based on certain qualitative features of these systems.  
    
    In the systems considered above, the configurations traversed form a discrete set, defined by locally stable configurations.
    The transitions associated with these traversals depend deterministically on the prior state and on the deformation applied to induce it.  The transitions are in general irreversible, so that a given configuration or state may have been reached from more than one possible prior state.  These features amount to saying that the evolution may be described as iterations of a map on a finite set. Upon iteration, such mappings generically yield convergence to a cycle \cite{FlajoletOdlyzko1990}, as is also observed in the physical examples above. 
    Here we develop a correspondence between cyclic annealing and the iteration of finite maps.  We can thus compare the behavior of the maps with that of physical systems.
    
    The convergence phenomena we aim to explain are well illustrated in the numerical studies of Regev {\em et al.}  and Fiocco {\em et al.} \cite{Regev2013,Regev2015,Fioccoetal2014}.  These system consists of a periodically continued box of a binary mixture with several thousand solid-like spheres.  These interact via a short range repulsive potential without friction.  The box is filled with enough spheres that they are held in position by the repulsions from their neighbors.  The annealing deformation consists of a shearing deformation of the box with a shear strain $\gamma$ that increases with time.  At every increment of strain the spheres are moved until local equilibrium is found.  The increments of strain are made small enough that limiting adiabatic behavior is reached. This is referred to as the AQS (athermal quasi-static) regime \cite{AQSMaloney2006}. 
    
    The increasing strain leads to mechanical instability.  The normal contact forces become aligned so that they can no longer support the applied shear stress, and the spheres move spontaneously without restoring force.  This motion is simulated as though dominated by viscosity, with no inertial forces.  The contact network evolves during this motion until a new equilibrium state---a new local minimum of potential energy---is reached .   In the simulations of amorphous packings these discontinuous motions or ``glitches" may involve few spheres or many.  After such a glitch has occurred, the shear strain again is increased as before, thus creating a sequence of glitches.  After some chosen maximum strain $\gamma_{\rm m}$ is reached,  the strain is reversed until the strain has decreased to $-\gamma_{\rm m}$.  Then $\gamma$ is reversed until it again reaches $+\gamma_{\rm m}$.  The strain continues to be varied through a number of such cycles.  Glitches occur throughout these cycles; increments of shear between glitches are generally small compared to $\gamma_{\rm m}$.  For systems of the size studied, one may clearly distinguish all glitches from numerical noise.
    
    During these annealing cycles, the glitches keep the same character but they differ in specifics: the displacements of the particles are different each time.  Moreover, these differences tend to decrease with succeeding cycles.  Ultimately, each glitch in the cycle produces the same displacements and occurs at the same strain as in previous cycles.  This constancy remains until the driving is changed in some way, \eg by a subsequent change of the maximum amplitude $\gamma_{\rm m}$.  Such changes disrupt the cycle and a transient of a number of cycles is required until convergence to a new cycle is achieved.  Once a given sequence of glitches is disrupted, it is generally not recovered.  The convergence process depends on the amplitude: larger amplitudes require more annealing cycles.  Further, there is generally a limiting amplitude $\gamma*$, called the yielding transition, above which convergence is impossible.  The convergence time is observed to diverge as $\gamma_m$ increases towards $\gamma^*$.   

    Above we have noted that cyclic annealing can lead a system from a very large set of states to a well-defined and small class of discrete states.  We have also observed that this process is deterministic; once any configuration is given and a direction of shear is specified, the next discontinuous change of state is uniquely determined.  This defines a map from an initial state to a succeeding state.  When extended to all possible states, such a map would dictate the sequence of transitions during increasing shear.  It  would be a complicated function of the disordered geometry of the system. Nevertheless, the convergence to a small subset of states when periodically driven suggest that these maps have common features that one can hope to capture by simpler models.  
    
    A similar convergence is an intrinsic property of discrete maps, without consideration of the mechanical origin of the mapping function.   Such maps occur widely in statistics and computer science \cite{ArneyBender1982,FlajoletOdlyzko1990,AnalyticCombinatorics}, and have also been studied within the context of spin glasses \cite{DerridaFlyvbjerg1987}.  Iter\-ating such maps produces convergence to a repeated limit cycle of one or more states, as with cyclic annealing in the sheared amorphous solids.  In the following, we aim to adapt such maps so as to describe annealing, with as little further restriction as possible.  
    
    Our approach of using random maps to model the irreversible and athermal dynamics of disordered systems has precedence. Within the context of 
    Kauffman's Boolean networks \cite{Kauffman1969} random maps have been considered for example in \cite{Bastolla1996,kulakowski1996relaxation,aldana2003boolean}. Much closer to our aims is the transition matrix (TM) approach of Fiocco {\em et al.} 
    \cite{Fiocco2015}, introduced in order to describe the athermal dynamics of a periodically sheared amorphous solid. In their approach an {\em ensemble} of random maps is constructed where each random map $P_{\gamma, \Delta \gamma}$ captures the transitions associated with a given strain change $\gamma \to \gamma + \Delta \gamma$ that is assumed to be small. The periodical shearing is implemented by a concatenation of these maps. Inspired by the TM approach, our work expands theirs by (i) noting that only two maps suffice to define the convergence process (ii) explicitly making connections with the theory of random maps and the available results in the literature, and (iii) using these connections to understand the relation between the structure of the maps and the resulting limiting behavior they produce.
    
    The paper is organized as follows. In the next section, Section \ref{sec:maps_scaling}, we recall the statistical features of arbitrary maps, also called random maps.  We focus on the number of limit cycles these maps contain and the average number of iterations needed to converge to them.  In these maps there is no notion of the sequence of strain induced by increasing or decreasing shear.  There is also no notion of repeated oscillation of the shear.  Thus in the following Section \ref{sec:annealing} we build a representation of annealing cycles using the discrete maps; it allows us to  distinguish between small-amplitude and large amplitude annealing. We then introduce in Section \ref{sec:samples} a succession of maps progressing from least restrictive to most restrictive, describing how each is constructed. In the subsequent Section \ref{sec:numerical}, we sample numerically from these maps and show how the convergence to a limit cycle depends on the number of states $\mathcal{N}$ in the set and on the annealing amplitude.  The convergence differs markedly from that seen in annealing simulations.  In the final section \ref{sec:discussion} we discuss possible reasons for these discrepancies.  There we point to further scope for improving the map representation of annealing.
	\begin{figure}[b]
    \includegraphics[width=\columnwidth]{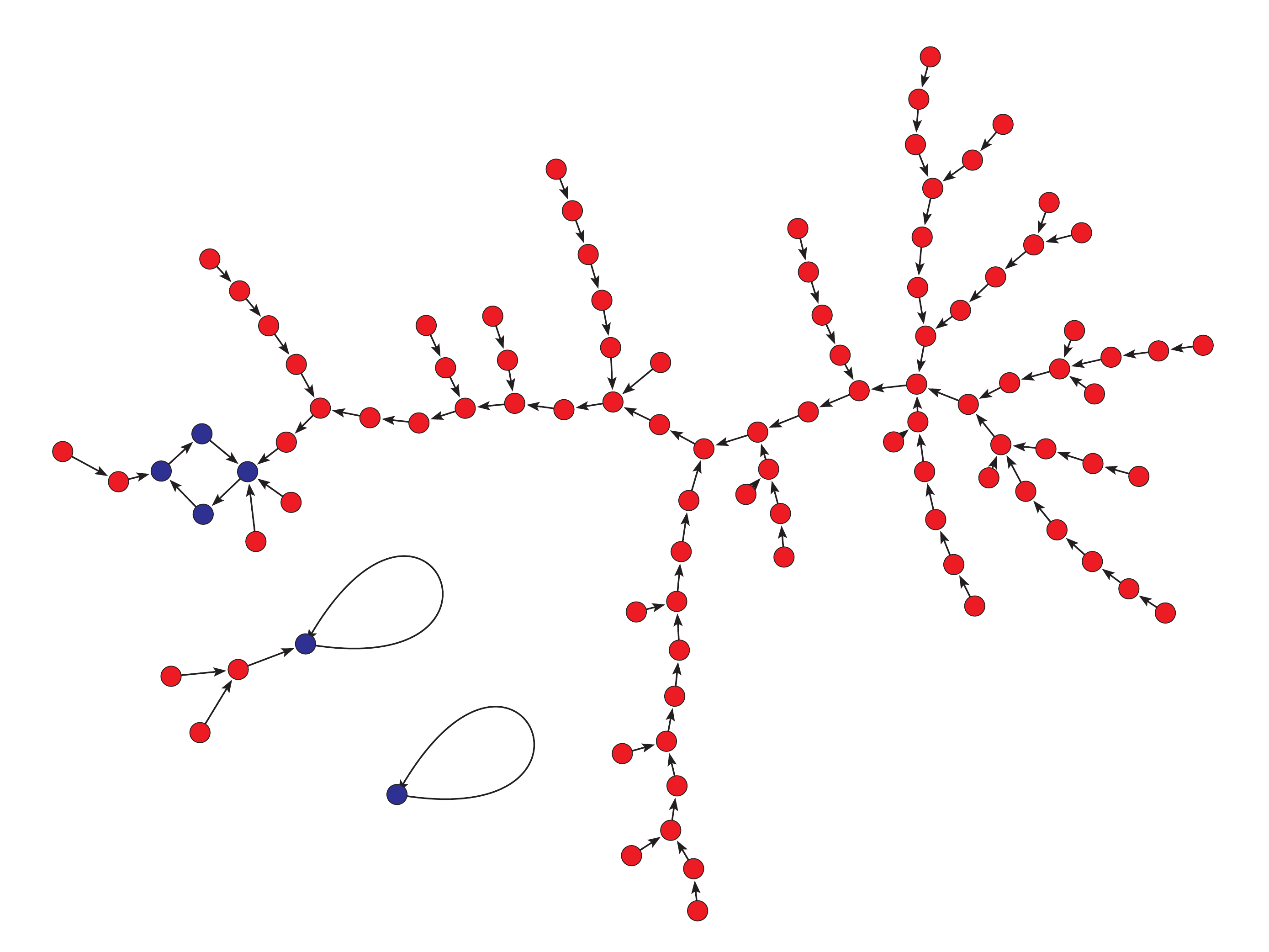} 
  \caption{Functional graph associated with a random map of $\mathcal{N} = 100$ elements. The functional graph consists of $n_c = 3$ components, each terminating in a limit-cycle. The lengths $\ell_c$ of the limit cycles are  $\ell_c = 4, 1, 1$ and the vertices of the limit cycles are colored in blue. Each blue cycle node is the root of a tree.  For three of the cycle nodes this tree is has only the blue root node; the other trees have 2, 3, and 87 additional red nodes.  The sizes of the components are $s_c = 95, 4, 1$.} 
  \label{fig:random_graph}
\end{figure}
			
    \section{Iterated discrete maps}\label{sec:maps_scaling}

 Given a set $S$ of $\mathcal{N}$ elements, a discrete map is a mapping from $S$ into itself. The properties of discrete maps have been well studied by group theorists \cite{Krohn:1965fk}.  In this background section we review the statistical properties explained \eg in Flajolet and Odlyzko \cite{FlajoletOdlyzko1990}, for later comparison with our results. 

It is convenient to think of a discrete map as a directed graph, the functional graph of the map, in which each element of $S$ is a node, and each node has a single directed edge pointing to some other node. Fig. \ref{fig:random_graph}  gives an example.  Evidently, the map can be such that a proper subset of $S$ may map into  itself and the functional graph may thus contain disjoint pieces. These pieces are known as ``components." Every node of such a graph has a sequence of images under the map, following each directed edge in turn. At some point in this sequence some element of $S$ must be revisited. The subsequent steps in the sequence must thence retrace previous steps, to return again and again to the first revisited element.  Given any initial element, this cycle is dictated by the map. Thus any given element of $S$ may only map to one cycle. The part of the sequence preceding this cycle is known as the ``tail". 
    
	\begin{figure*}[t!]
\centering
    \includegraphics[width=\textwidth]{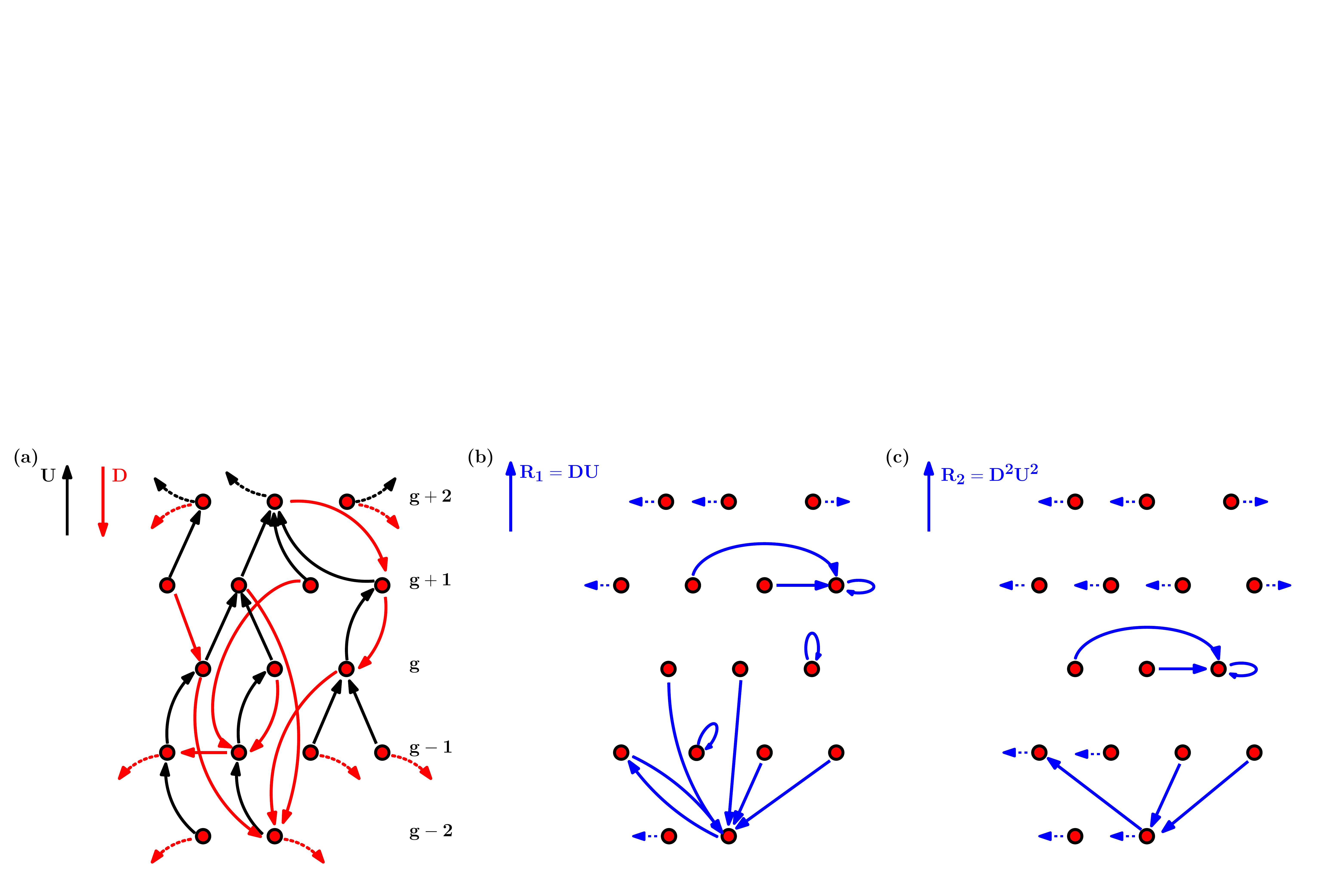} 
\caption{Functional graphs for iterated $\Up$ and $\Dn$ maps.  a) Graph detail for a $\Up$ tree and a compatible $\Dn$ tree.  Nodes (red dots) are arranged in rows having a common distance $g$ from the $\Up$ root.  The $\Up$ mappings are shown as black arrows; $\Dn$ mappings are shown as red arrows.  Dotted edges go to nodes not shown.  These maps are ``generation compatible": every $\Dn$ arrow connects from a node to another below it.
  b) Graph detail of the map $\mathbf{R}_1 \definedas \Dn \Up $ showing three fixed points and one 3-cycle.  c) Graph detail of the map $\mathbf{R}_2\definedas \Dn\Dn~ \Up \Up$ showing a single fixed point.}
  \label{fig:sample_map_constructions}
\end{figure*}

    Just as every element leads to a cycle,  two elements in the same component must lead to the same cycle, as we now argue. Being in the same component means  that the two elements are connected via the functional graph.  That is, following the graph from each of the two elements must lead to one common element.  Subsequent elements after the first common one are necessarily also common.  This includes the cycles of both elements.  Thus two elements in the same component necessarily share a cycle.  Thus every component has exactly one cycle.  The end of the tail of a given element is an element of this cycle.  The tails of numerous elements may meet at this end point.  The subgraph of these elements may contain no cycles; it is thus a tree graph whose root is the end point.  The union of all these roots is the cycle.  

    We next consider an arbitrary element and the  $\ell_t$ elements of its tail leading to its cycle.  Any element in this tail may be joined by tails from other elements.  
    We now consider a second arbitrary element.  
    If the tail of this element does not lead to the same cycle, it must therefore  belong to another component.  We use this fact below to estimate the number of components.

    To summarize, every map on $\mathcal{N}$ elements consists of some number $n_c$ of disjoint components, each of which contains a single cycle with some $\ell_c$ elements.  To each cycle there are attached disjoint directed trees, each rooted in an element of the cycle.  A given element in such a tree is joined to its cycle by some $\ell_t$ tail elements. The number of elements in a component
    is denoted by $s_c$, the size of the component. 
    The functional graph of a typical random map is illustrated in Fig \ref{fig:random_graph}.

    We next consider the ensemble of discrete maps of a set of $\mathcal{N}$ elements into itself. We call such a map a {\em simple random map} in  
    order to distinguish it from the map ensembles to follow. Using generating-function techniques \cite{FlajoletOdlyzko1990}, it is possible to work out how the expectation values of the features defined above vary asymptotically with $\mathcal{N}$:
    \begin{align}
      n_c &= \frac{1}{2} \ln \mathcal{N} + \frac{1}{2} \ln 2 + \gamma, \label{eqn:nc} \\
      s_c &= \frac{2}{3}\mathcal{N}, \label{eqn:sc} \\
      \ell_c &= \ell_t = \sqrt{\frac{\pi \mathcal{N}}{8}}, \label{eqn:ellc}
    \end{align}
    where 
    $\gamma$ is the Euler-Mascheroni constant \footnote{Since our discussion is mostly quantitative, we shall use the same letter to refer to an observable and its ensemble average, assuming that the meaning will be clear from the context. Thus in \eqref{eqn:nc} $n_c$ refers to the number of components of a map averaged over the ensemble of random maps.}. The statistics for $\ell_c, \ell_t,$ and $s_c$ are formed by  averaging also over the nodes. For example $s_c$ is the size of the component that a node belongs to, where both the map and the node are selected at random and uniformly. Likewise, $\ell_c$ is the expected length of the cycle associated with this component {\em etc.} Further details can be found for example in  \cite{Kruskal54, FlajoletOdlyzko1990, AnalyticCombinatorics}, and references therein.  
			
    It is instructive to obtain the large-$\mathcal{N}$ scaling behavior of these estimates from simple arguments. To illustrate this, we note that for random maps the ensemble for $\mathcal{N}+1$ elements may be constructed inductively from its $\mathcal{N}$-element counterpart.

    The average number of components in a random map $n_c(\mathcal{N})$ may be estimated in this way \cite{Kruskal54}.  We consider the average number of new components $\Delta n_c(\mathcal{N})$ that appear when one new element is added to $\mathcal{N}$.  The new element must map to one of the existing elements or to itself.  Only the latter choice creates a new component.  And this choice has a probability $1/(\mathcal{N}+1)$.  Adding these increments for integers up to $\mathcal{N}$ yields $n_c(\mathcal{N}) \goesas \ln \mathcal{N}$.  
    Likewise, the average number of elements in a component $s_c$ scales as $\mathcal{N}/\ln \mathcal{N}$.  

    If an element of a map is chosen at random and its successive images are recorded, there must come a point where a prior element is repeated.  The sequence from this prior element to the current one must then repeat cyclically, as noted above.  The average length of a sequence to the first repeated element is called its {\em six-length}  $\ell_6$; all of its elements are necessarily mutually distinct.  The probability that the {\it first} step in this sequence happens to be a repetition is $1/\mathcal{N}$ since the first node's image must be itself.  For large $\mathcal{N}$, the probability that the second step is a repetition, becomes $2/\mathcal{N}$: the second node's image may be either the first node or itself.   Similarly, the first repetition occurs at the $k$'th step with probability $k/\mathcal{N}$ provided  $k \muchlessthan \mathcal{N}$.  Thus the cumulative probability of a repetition after $k$ steps is of order $k^2/\mathcal{N}$.  This probability grows to the order of unity when $k^2 \goesas \mathcal{N}$.  At this point $k$ is of the order $\ell_6$, so that $\ell_6{}^2 \goesas \mathcal{N}$.  The six-length $\ell_6$ consists of two parts: the ``tail" sequence, of length $\ell_t$ preceding the repeated element and the cyclic sequence of length $\ell_c$ following it.  Since given a six-length sequence, the starting position of the cyclic segment is equally likely to be anywhere on it, the expectation value of the tail and cycle lengths must be the same.  Thus $\ell_t \sim \ell_c \sim \ell_6 \sim \sqrt{\mathcal{N}}$.

    It is useful to introduce some notation. Let $\mathbf{P}$ and $\mathbf{Q}$ be two maps that map a set of $\mathcal{N}$ elements into itself. Then the action of the {\em composite map} $\mathbf{P}\mathbf{Q}$ on some element $A$ is obtained by mapping $A$ first according to $\mathbf{Q}$, the result of which 
    we denote by $\mathbf{Q}A$, and then applying $\mathbf{P}$ to it, to obtain the element $\mathbf{P}(\mathbf{Q}A) \equiv \mathbf{P}\mathbf{Q}A$, by associativity. Observe that we apply a sequence of maps from right to left. 
    An {\em iteration} is a composition of a map with itself, \eg $\mathbf{P}\mathbf{P} \ldots \mathbf{P}$ which we write as $\mathbf{P}^n$, with $n$ indicating the number of iterations. 

    A property of maps that will prove important below is the {\em contraction factor}: 
    a given element $A$ maps to exactly one element, but more than one element may map to $A$.  An element may also have no such inputs.  The contraction factor $z$ of a map  is the average number of inputs to an element with at least one input.  Thus the number of elements contracts by a factor $z$ upon iteration.  
    The fraction of the elements with no inputs is given in terms of $z$ as  $1 - z^{-1}$. Arney and Bender \cite{ArneyBender1982}  report results  for a variety of cases where the random maps are further constrained by the specification of the contraction factor $z$. In particular, they find that the
    large $\mathcal{N}$ scaling of the number of components $n_c$ and their size $s_c$ remains unaltered, given by \eqref{eqn:nc} and \eqref{eqn:sc}, while the cycle and transient lengths now become
    \begin{equation}
      \ell_c = \ell_t = \sqrt{\frac{\pi \mathcal{N}}{8\lambda}},
    \end{equation}
    with $\lambda$ depending on $z$ in a known way. This equation is identical to Eq. \eqref{eqn:ellc}, with $\cal N$ replaced by $\cal N/\lambda$.   Thus the scaling with $\cal N$ is unaltered.

    \section{representation of annealing} \label{sec:annealing}
    The discrete maps defined above represent two features of the cyclic annealing process we wish to treat.  They represent the discreteness of the sequence of states and the determinism of the transitions between them.  Here we make further restrictions on the maps to enable a more faithful description of annealing.  This description raises potential confusion between two notions of cycles, which we now take pains to clarify.  In the preceding section we noted that all finite maps when iterated terminate in limit cycles.  These are intrinsic to the map in question and have nothing to do with cyclic driving.  The physical annealing processes of \eg a granular packing discussed in the Introduction are not intrinsic to the system but are imposed externally through the periodic driving.  
    We will call the repetitions constituting this periodic driving ``annealing cycles."  Our goal is thus to see how annealing cycles can be represented by maps and to compare the behavior of the maps with that of the annealing simulations. Specifically, we wish to reproduce the physical phenomenon in which a repetition of annealing cycles leads to a cyclic repetition of configurations, as described above.

    We have noted that the increasing of the imposed strain causes a sequence of discrete changes of state---the glitches defined above.  Our mapping scheme gives us no way to infer the strains at which a glitch occurs.  Still, we may use the glitches as a proxy for the transition from one strain state of the sample to another.  Thus for any packing in a particular configuration $A$ that is stable under a given strain, the configuration $B$ obtained by {\em adiabatically} increasing the strain through a single glitch is a deterministic function of $A$, leading to a new stable state that we denote as $\Up A$.  The ensemble of successors of all configurations $A$ under increasing strain defines the map $\Up$.  Likewise, if the configuration $A$ is subjected to a {\em de}creasing strain through a single glitch, this results in a new configuration $C$ and defines an analogous map  $\Dn$. We write this transition as $\Dn A = C$.  

    Conventional annealing is done by monitoring the driving parameter (\eg the strain $\gamma$) at which a given configuration shifts to another as one proceeds through a fixed range of $\gamma$ values. Our proxy for this procedure is to repeatedly increase the strain through $p$ glitches and then decrease it through $q$ glitches. 
    The map for a complete annealing cycle is thus given by  $\Dn^q \Up^p$.  
    This map need not correspond to an explicit range of $\gamma$: our driving is oscillatory but $\gamma_{\rm max}$ and $\gamma_{\rm min}$ can change from driving period to the next. Still, our procedure does prescribe a form of cyclic annealing. 
    If repeating this process leads to a limiting cycle of configurations, then that cycle necessarily occurs for some fixed limits $\gamma_{\rm min}$ and $\gamma_{\rm max}$.  
    
    For simplicity we shall consider an annealing cycle with $p$ upward glitches followed by $p$ downward, ones. In terms of our maps $\Up$ and $\Dn$ this 
    corresponds to iterating $\Up$  $p$ times and composing the result with $\Dn$ applied $p$ times to form the ``annealing cycle map"  
    \begin{equation}
      \mathbf{R}_p = \Dn^p \Up^p. \label{eqn:Rpmap}
    \end{equation}
    A pattern of glitches that repeats with every annealing cycle corresponds to a fixed-point of $\mathbf{R}_p$,  mapping the state at the beginning of the annealing cycle into itself. We shall call this type of response {\em synchronous response}. 
    Larger limit cycles of $\mathbf{R}_p$ would correspond to {\em sub-synchronous response}  \cite{Deutschetal2003}, \ie repetition of the states after two or more annealing cycles, as observed in some simulations and experiments \cite{Regev2013, Lavrentovich2017}. 
    
    \paragraph*{Functional graphs of $\Up$ and $\Dn$; partial ordering.}
    
The $\Up$ map is defined by increasing strain; this constrains the structure of its functional graph.  Indeed, the strain $\gamma^+(A)$ at which a given state $A$ undergoes an up transition must be lower than that of its image, \ie $\gamma^+(\Up A)$. Thus successive up transitions, must have strictly increasing $\gamma^+$ values.  Now, it is possible for a given state to be repeated at some larger shear value.  This means that a state may have more than one external increasing strain that produces its glitch.  Once such a state is encountered, the subsequent $\Up$ maps must follow a cycle.  Thus cycling of the states (or a fixed point) would occur under increasing strain, without the need to do external cycling. This simple behavior does not  appear to occur for the small shear amplitudes (less than $\gamma^*$) of interest here.  Thus for our purpose we may suppose that no state $A$ can be repeated upon successive $\Up$ maps 
\footnote{Such an assumption is certainly valid for analogous systems,  such as spin-glasses or elastic manifolds in random media. However in these cases the adiabatically increasing variable was a force, {\em e.g.} an external magnetic or electric field. The sheared systems we consider here are driven differently. The adiabatically controlled variable is not a force but a constraint, the geometrically-imposed strain.}.    
The functional graph must thus be a tree or a group of disjoint trees or forest.  Such {\em acyclic} maps are necessarily partially ordered, since each element has a fixed number of steps  to the root of its tree. Thus there is a fixed order of precedence in the path from a given state to that root. The $\Dn$ map is also acyclic, tree-like  and partially ordered by the same reasoning.  Since annealing convergence happens for mild strain cycles far below the yielding strain, we suppose that the depths of these trees are large compared to the number of glitches $p$ in an annealing cycle. 

The partial ordering property constrains $\Up$ and $\Dn$ individually but gives no constraining relationship between $\Up$ and $\Dn$.  However, in order for  the  pair of maps $(\Up, \Dn)$ to correspond to annealing, it should obey some additional conditions, which we now address.

    \paragraph*{Compatibility}  Every state $A$ in the physical system has a stable interval of strain bounded by an upper and a lower strain denoted $\gamma^+(A)$ and $\gamma^-(A)$.  The spread between the $\gamma^-$ of a state and its $\gamma^+$ is a natural measure of its stability.  If this spread is much narrower than the range of strains $\gamma$ imposed during an annealing cycle, one may say that the states have little stability.  One can gauge the degree of stability by considering an arbitrary state $A$ that is stable for some strain $\gamma$, and then increase or decrease $\gamma$ until a glitch occurs.  The initial $\gamma$ necessarily lies between $\gamma^-(A)$ and $\gamma^+(A)$.  The necessary change in $\gamma$ is then a rough gauge of the spread $\gamma^+ - \gamma^-$ of the that state, which we denote by $\delta \gamma(A)$.   In the rich final states we seek to explain, there are many glitches in the annealing cycle.  Accordingly, we will focus on systems whose $\delta \gamma$ is small on the scale of the annealing range. Following the above reasoning, we shall suppose $\delta \gamma$ to be comparable to the typical interval of $\gamma$ between glitches as one traverses the annealing cycle.  

    We now consider the successor to $A$ upon increasing the strain to $\gamma^+(A)$, \ie $\Up A$.  It can happen that the lower strain $\gamma^-(\Up A)$ is higher than $\gamma^-(A)$.  In that case, the successor under decreasing strain, \viz $\Dn A$, cannot be $\Up A$, since $\gamma^-(D A)$ must be smaller than $\gamma^-(A)$.  We say that this choice of $\Dn A$ is {\em incompatible} with $\Up A$.  

    When the upper and lower strains have a small spread of strains $\delta \gamma$, this puts a compatibility constraint on the pair of maps $(\Up,\Dn)$.  Each glitch encountered upon increasing strain from a state $A$ increases the $\gamma^+$ by an amount of order $\delta \gamma$.  By our reasoning above, it also increases its $\gamma^-$ by a comparable amount.  On average then, the new $\gamma^-(\Up A)$ lies above $\gamma^-(A)$.  Successive upward glitches from $A$ steadily raise the typical values of both $\gamma^+$ and $\gamma^-$.  Thus they steadily diminish the probability that the new $\gamma^-(\Up^n A)$ lies below the original $\gamma^-(A)$.  Each successor is thus less likely to be compatible with $A$.

    Now take the first glitch encountered upon {\em de}creasing the strain from state $A$ and consider $\Dn A$.  The $\gamma^-$ of this state \ie $\gamma^-(\Dn A)$ is necessarily below $\gamma^-(A)$.  According to the last paragraph, the $\Up$ successors of $A$ have limited likelihood of being compatible with $\Dn A$. Furthermore, successors $\Dn \Dn A$, {\em etc.} have progressively less likelihood of being compatible. 

    Evidently the requirement of compatibility limits the $\Dn$ maps  that are admissible for a given $\Up$ map.   Clearly, one cannot decide whether a given $\Dn$ map is compatible without knowledge of the $\gamma$ range associated with each state $A$.  Still as argued above, the compatibility condition influences the possible $\Dn$ and $\Up$ successors of a state $A$: namely states in the $\Dn$ tail of $A$ are unlikely to be in the $\Up$ tail of the same $A$.  We may eliminate these unlikely events entirely without knowledge of the $\gamma$ ranges.  Accordingly, we denote any ($\Up$, $\Dn$) pair where the $\Up$ and $\Dn$ tails of every $A$ are disjoint as a {\em tail-compatible} pair.   Below we shall consider the effect of tail compatibility and stronger conditions on the convergence properties of maps.  

	    \paragraph*{Simplifying restrictions} For the sake of simplicity, we will constrain the $\Up$ and $\Dn$ maps even further.  First, as justified above, we shall suppose that both $\Up$ and $\Dn$ are rooted trees.  This means that each state inherits a label providing the number of $\Up$ steps to the $\Up$ root. We denote this label as the $\Up$ {\em generation} of the state.  Each state $A$ has an analogous $\Dn$ generation.  A step in the $\Up$ direction necessarily decreases the $\Up$ generation of 
    $A$ by one.  This step need not change its $\Dn$-generation. However we shall ensure tail-compatibility of $\Up$ and $\Dn$ by restricting the $\Dn$ generation of 
    $\Up A$:  we require that under a step of $\Up$ or $\Dn$, the two generations may not both increase or decrease. We call this condition {\em generation compatibility} and note that it suffices to achieve tail compatibility as defined above.  We show an example of two generation-compatible maps in Fig. \ref{fig:sample_map_constructions}.

    These restrictions, aimed at representing realistic conditions of cyclic annealing, are minimal and they constitute a {\em null-model} that captures the irreversible transitions under adiabatic increases and decreases of the driving parameter. Our aim therefore is to explore numerically (i) whether these restrictions have a potentially large effect on the scaling behavior when compared with that of generic random maps, and (ii) how well these restrictions reproduce the observed behavior seen in cyclic annealing experiments.  
				
			\begin{figure*}[t!]
\centering
    \includegraphics[width=\textwidth]{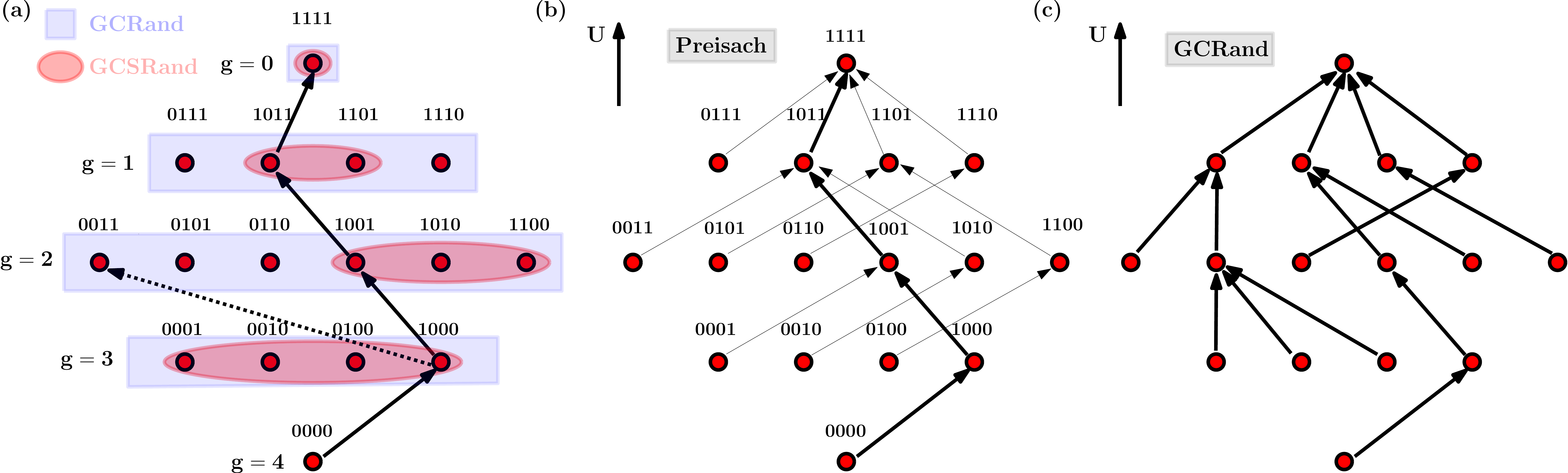} 
\caption{Functional graphs for {\tt GCRand} {\tt GCSRand} and {\tt Preisach} maps with $2^4$ nodes compared.  Nodes shown as red dots are arranged by generation as in Fig. \ref{fig:sample_map_constructions}.  Spin labels described in the text are shown above each node.  Thus each row contains the same number of 1 spins.  Within a row the labels are arranged in numerical order.  In all three panels, the ``backbone" $\Up$ edges leading from the $\Dn$ root to the $\Up$ root are indicated.  a) contrasts {\tt GCRand} and {\tt GCSRand} maps on these nodes.  For {\tt GCRand} maps, each node may map to any node in the row above. For {\tt GCSRand} the choices are restricted as indicated by the red shading.  Starting from the $g=4$ node, all $g=3$ options are allowed (since all have a single 1 spin).  Given the choice made for this mapping, the next mapping to the $g=2$ row may only be one of the red-shaded nodes in that row.  Thus the dashed arrow is not allowed in {\tt GCSRand}.  b) shows the {\tt Preisach} map with the same backbone as in a).  Here, the ordering of the Preisach $\gamma^+$'s is dictated by the backbone.  Since this ordering determines the entire $\Up$ map, the rest of the map, shown in light arrows, is determined.  c) shows a general {\tt GCRand} map with the same backbone.}
  \label{fig:GCandGCS}
\end{figure*}

    \section{Sample maps}\label{sec:samples} We consider four types of map-pairs $(\Up,\Dn)$, progressing from least restrictive to most restrictive. 
    The least restrictive takes $\Up$ and $\Dn$ to be a pair of simple random maps in the sense of \cite{FlajoletOdlyzko1990}, and we denote this by {\tt Rand}.  We next add the restriction of compatibility and then impose increasingly stronger additional constraints, leading in turn to the map-classes, {\tt GCRand}, {\tt GCSRand}, 
    and finally {\tt Preisach}. We describe their constructions next.

    \paragraph*{{\tt Rand}} We present this example to show the effect of composing two maps without further restrictions. Here $\Up$ and $\Dn$ are two independently constructed simple random maps, each mapping a set of $\mathcal{N}$ elements into itself, as described in Section \ref{sec:maps_scaling}. We are thus ignoring all 
    restrictions considered in Section \ref{sec:annealing}.       
    Let $\Omega_{\tt Rand}$ be the set of all $(\Up,\Dn)$ map-pairs that can be constructed in this manner. Thus $\Omega_{\tt Rand}$ has 
    $\mathcal{N}^{2\mathcal{N}}$ elements \footnote{Note that the number of distinct maps $\mathbf{R}_p$  representing any annealing cycle can be no greater than $\mathcal{N}^\mathcal{N}$.  Thus there must be many choices of ($\Up$, $\Dn$) that yield the same $\mathbf{R}_p$. }. 
    
    \paragraph*{{\tt GCRand}} These ``generation-compatible random" maps obey the restrictions introduced in Section \ref{sec:annealing}. They are intended to show how much these restrictions affect the scaling properties seen in {\tt Rand}. Thus we now take $\Up$ and $\Dn$ to be  trees. We implement a strong form of generation compatibility as shown in Fig.  \ref{fig:GCandGCS}: Given a node $A$ with $\Up$-generation $g$, we require that $\Up A$ and $\Dn A$ are nodes of 
    generation $g - 1$ and  $g + 1$, respectively. Denoting by $L$ the total number of $\Up$-generations, it follows that a state having $\Up$-generation 
    $g$ has simultaneously $\Dn$-generation $L-g$ \footnote{We have also tried out variations of {\tt GCRand} where the tail-compatibility between $\Up$ and $\Dn$ is enforced in a less restrictive way. For example, under $\Dn$ we also allowed for the $\Up$ generation to stay constant. We found qualitatively similar results. 
    }.
    We  further let $\mathcal{N}(g,L)$ be the number of nodes in generation $g$. 
    Equating the corresponding generations of $\Up$ and $\Dn$, we require that $\mathcal{N}(g, L) = \mathcal{N}(L-g, L)$\footnote{The requirement that $\mathcal{N}(g, L) = \mathcal{N}(L-g, L)$ is also  motivated by the observation that 
      $\Dn$ is related to $\Up$ by reversal of the direction of driving and their equivalence.}. In particular, we will choose 
    \begin{equation}
      \mathcal{N}(g,L) = \left ( \begin{array}{c} L \\ g \end{array} \right ), 
      \label{eqn:binom}
    \end{equation}
    to be the binomial coefficients, so that $\mathcal{N} = 2^L$. Clearly, this strict generation compatibility entails the looser generation compatibility defined above and the latter in turn assures the tail compatibility property related to physical constraints on packings. 
    
    Such maps have a natural interpretation in terms of a specific configuration set, namely the $2^L$ configurations of $L$ two-state ``spins" denoted ``up" and ``down."  Here the generation corresponds to the number of down spins. The $\Up$ map is one that decreases the number of down spins by one.  The $\Dn$ map is one that increases that number by one.  Evidently, no annealing cycle may have more than $2L$ glitches.

    The map $\Up$ of {\tt GCRand} is then constructed as follows. We first designate a set of nodes for each generation $g$ from $0$ to some $L$.  The number of nodes assigned to generation $g$ is the $\mathcal{N}(g, L)$ of Eq.~\eqref{eqn:binom}.  Then for each node $A$ of generation $g$, we assign for $\Up A$ a node picked uniformly and at random among those in generation $g-1$. The map $\Dn$ is generated independently and in a similar manner. We denote by $\Omega_{\tt GCRand}$ the set of all pairs $(\Up,\Dn)$ of maps that can be constructed in this way. $\Omega_{\tt GCRand}$ has $\prod_{g=1}^L\mathcal{N}(g,L)^{2\mathcal{N}(g-1,L)}$ elements. 
    
    \paragraph*{{\tt GCSRand}} These ``generation-compatible single random" maps are a subset of $\Omega_{\tt GCRand}$. We think of the states as spin configurations, and  
    as in {\tt GCRand}, $\Up A$ is a configuration with one more up-spin than $A$. However, in {\tt GCSRand} $\Up A$ is chosen among configurations that differ from $A$ by a {\em single} spin changed from down to up.  Likewise, $\Dn A$ is the $A$ configuration with one up-spin changed to a down-spin. Evidently, the number of 
    elements in $\Omega_{\tt GCSRand}$ is given by $\prod_{g=1}^L g^{2\mathcal{N}(g-1,L)}$. 

    \paragraph*{{\tt Preisach}} The Preisach model \cite{Preisach1935} is a spin model of annealing for which the pair of maps $(\Up,\Dn)$ can be constructed as well.  Here each binary spin  $i$ evolves based on two shear thresholds: a smaller one  $\gamma_i^-$ and a larger one $\gamma_i^+$. We further require $\gamma_i^- < \gamma_j^+$ for all $i, j$ \footnote{This can be achieved for example by requiring as in \cite{Barker1983} that   
    $\gamma^-_i < 0 < \gamma^+_i$ for all $i$.}. In this case, only the rank orderings of  $\gamma_i^+$ and $\gamma_i^-$ suffice to determine $\Up$ and $\Dn$.
 Moreover, for each of the $2^L$ configurations, there exists a range of values $\gamma$ over which none of the spins flip.  Whenever the applied shear $\gamma$ increases through $\gamma_i^+$ the corresponding spin $i$ becomes or remains 1.  Thus increasing $\gamma$ causes a succession of spins to increase, one at a time.  This defines a $\Up$ map from any given spin state that proceeds by single spin flips.  Specifically, $\Up A$  is the state that one obtains from $A$ by raising the 0 (down)  spin having the smallest $\gamma_i^+$.  Upon {\em decreasing} $\gamma$, every spin $i$ becomes 0 when $\gamma$ decreases through $\gamma_i^-$.  A  $\Dn$ map is similarly defined.  Given any spin state: $\Dn A$ is the state in which the 1 (up) spin of $A$  with the largest $\gamma_i^-$ is lowered.  Both maps proceed by single spin flips only.  Thus the Preisach model obeys the rules of the {\tt GCSRand} class, and is a subset of that class.       
\begin{figure*}[t!]
\centering
    \includegraphics[width=\textwidth]{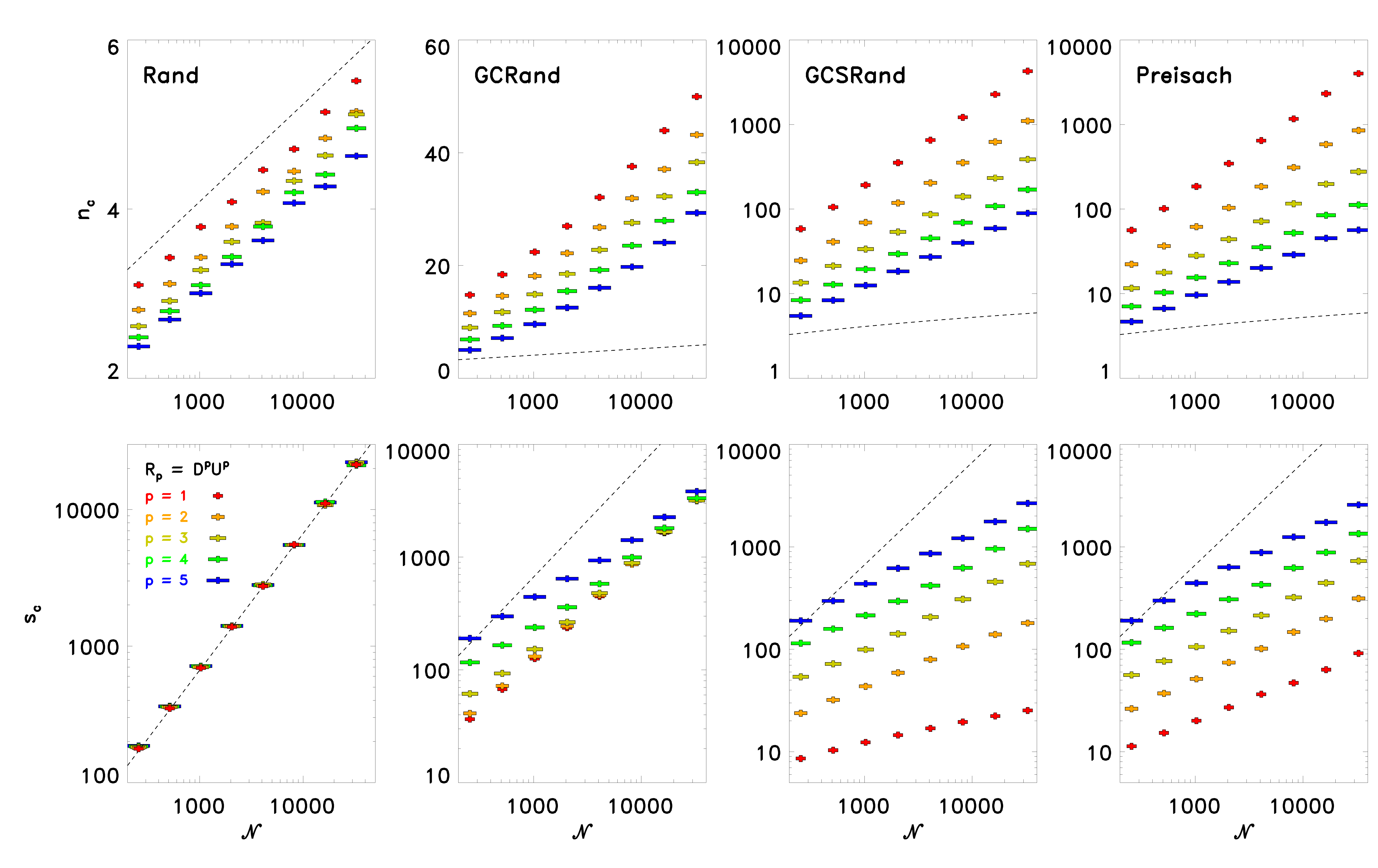} 
  \caption{Sample-averaged statistics of $\mathbf{R}_p = \Dn^p \Up^p$ for the random map classes {\tt Rand}, {\tt GCRand}, {\tt GCSRand} and {\tt Preisach} (from left to right column), introduced in Section \ref{sec:annealing}. The first and second rows show the average number of components $n_c$ and  component sizes $s_c$ versus system size $\mathcal{N}$.
  The color and width of the symbols correspond to the different values of $p$, as indicated in the lower left plot. The black dashed lines are the predictions for the simple random map, \eqref{eqn:nc} and  \eqref{eqn:sc}.  } 
  \label{fig:num_data}
\end{figure*}
Denoting by $\Omega_{\tt Preisach}$ the set of all Preisach map pairs $(\Up,\Dn)$, we see that   $\Omega_{\tt Preisach}$ has $(L!)^2$ elements. 
    
    Note in particular that $\Omega_{\tt Preisach} \subset \Omega_{\tt GCSRand} $, \ie every {\tt Preisach} necessarily is also a {\tt GCSRand} map. The converse is not true in general, as follows immediately from a comparison of the number of elements that these two sets contain.

    Since each class of maps was defined by making restrictions on the previous class, the sets $\Omega$ are related by
    \begin{equation}
      \Omega_{\tt Preisach} \subset \Omega_{\tt GCSRand} \subset \Omega_{\tt GCRand} \subset \Omega_{\tt Rand}, 
    \end{equation}
as illustrated in Fig. \ref{fig:GCandGCS}. 
    Moreover, the elements in one subset constitute an asymptotically vanishingly small fraction of the elements in the set containing it, as is readily shown. Note that conceptually we can think of these maps 
    as mean-field models, since they do not contain any reference to an underlying spatial structure. Instead, the mapping pattern is largely determined by assigning nodes to classes, prescribing where members of these classes can map to and then selecting the map images randomly from the possible choices.   
    
    Now that the $\Up$ and $\Dn$ maps have been defined for these different models, we may readily determine the effect of an annealing cycle $\mathbf{R}_p = \Dn^p \Up^p$ as described in Section \ref{sec:annealing}.  Once an $\mathbf{R}_p$ is constructed, we may readily analyze the convergence to limit cycles as defined in Section \ref{sec:annealing}.  
    
    This response is especially simple in the Preisach model, due to its {\em return-point-memory} (RPM) feature \cite{Barker1983}, which the other classes of maps lack in general.  The ``return point" $\gamma_m$ of an annealing cycle is the point at which the direction of shearing reverses.  It is thus the point at which $\gamma$ is algebraically maximal or minimal. A well-known route to RPM is via the no-passing property (NP) \cite{NoPassing}, which is a dynamic constraint enforcing the preservation of some partial ordering of the states. NP implies RPM \cite{Sethna93}.  Return point memory via NP implies that repeated iterations with a given return point $\gamma_m$ revisit the same configuration whenever $\gamma$ returns to $\gamma_m$ \cite{Sethna93}.  Thus convergence is always achieved in one annealing cycle, the first time $\gamma_m$ is revisited. Moreover, under NP the response is synchronous, \ie $\ell_c = 1$ \cite{munganterzi2018}. It can be shown that the dynamics of the Preisach model has the NP property \cite{TerziMungan2019}.  
    
    The RPM property restricts the {\em hysteresis loop} of the system, \ie the plot of total spin as a function of strain.  In a system with RPM, these loops are nested:  if a large-amplitude strain cycle is followed by a smaller amplitude one, the smaller loop lies within the larger one.   The general structure of the functional graphs associated with map pairs $(\Up,\Dn)$ when RPM is present, has been worked out in \cite{munganterzi2018}.  
    
   	Evidently the RPM property requires immediate convergence to a fixed sequence of states from the first annealing cycle onwards once the range of shear $\gamma$ has been established.  This convergence is less immediate for the $\mathbf{R}_p$ annealing cycles explored below.  These $\mathbf{R}_p$ cycles do not entail a fixed range of strain until the limiting behavior is reached.  Thus the convergence length $\ell_t$ need not be reached in a single iteration of $\mathbf{R}_p$.  
        
    \section{numerical sampling}\label{sec:numerical} 

    In this section we present the results of our numerical sampling from the maps introduced in Section \ref{sec:samples}. 
    Our ultimate goal is to compare qualitatively with the annealing experiments the following quantities: (i) the length of the transients before limit cycles set in, (ii) the length of the resulting limit cycles, and (iii) the dependence of these quantities on the map properties. We explore how each of the classes of $(\Up, \Dn)$ map pairs converges to its limiting behavior under cyclic annealing in the limit of large system sizes $\mathcal{N} \definedas 2^L$ and how this convergence depends on $p$. We start out with a description of the sampling. We then briefly summarize our main findings before presenting these in detail.
				    
    For a given $\mathcal{N}$, we examine annealing cycles of the form $\mathbf{R}_p = \Dn^p \Up^p$ for amplitudes $p = $ 1---5.  We examine $L$ ranging from $8$ to $15$ so that $\mathcal{N}$ 
    ranges from $256$ to $32768$, spanning a little more than two decades. For each size $\mathcal{N}$ and type of map construction, we generate $\mathcal{R}_L$ random realizations, where $\mathcal{R}_L = 1000 ~ (2^{13-L})$. We then identify the different components and calculate  $n_c$, $\ell_c$, $\ell_t$ and $s_c$, as defined in Section \ref{sec:maps_scaling} for each map and perform averages over the number of realizations. 
    
    Let us summarize our main numerical findings next. As we go from {\tt Rand} to {\tt Preisach} by gradually imposing more restrictions on the map, the resulting convergence behavior changes. The map class {\tt Rand} behaves like the simple iterated random map, whose scaling properties are given by \eqref{eqn:nc} -- \eqref{eqn:ellc}. In {\tt GCRand} with generation compabibility, the number of components is increased. But in both {\tt Rand} and {\tt GCRand} the components structure is dominated by a few large components whose size scales linearly with the system size. Most of the nodes are trapped into limit cycles with long periods with respect to the annealing cycles giving rise to sub-synchronous response \ie $\ell_c > 1$, as defined in Section \ref{sec:annealing}. Typical cycle lengths grow as $\sqrt{\mathcal{N}}$. Adding further restrictions in going from {\tt GCRand} to {\tt GCSRand}, we observe that the component structure where a few macroscopic components contain most of the state is broken up now. Consequently, in {\tt GCSRand} transient lengths and lengths of limit-cycles are substantially reduced. Sub-synchronous response is still present, but $\ell_c$ tends to one with decreasing $p$. Finally, in the {\rm Preisach} family of maps, we find $\ell_c \equiv 1$ and $\ell_t \leq 2$. Both of these are a consequence of the no-passing property, which in turn implies the RPM property, as discussed above. Thus the classes of maps we consider here span a broad range of response. We now turn to a detailed discussion of these results. We will relate  these findings  
    to the experimental annealing results in the Discussion Section \ref{sec:discussion}. 
    
    \subsection{Component size and number}  We start out by looking at the component structure emerging from the different classes of maps. 
    In Fig.~\ref{fig:num_data} we show our results for the number of components $n_c$ and their sizes $s_c$ for the four types of maps considered. We have plotted the sample averaged statistics of the maps $\mathbf{R}_p =  \Dn^p \Up^p$, against system size $\mathcal{N}$ for different powers $p$. 
    From left to right, each column corresponds to one of the four random maps: {\tt Rand}, {\tt GCRand}, {\tt GCSRand} and {\tt Preisach}, respectively, with  the first and second rows showing the behavior of $n_c$ and $s_c$. The color and width of the plotting symbols represent $p$ and the legend is given in the inset of the bottom-left plot. The black dashed lines in each plot are the predictions for the simple random map, \eqref{eqn:nc} and \eqref{eqn:sc}.  
				
\begin{figure*}[t!]
\centering
    \includegraphics[width=\textwidth]{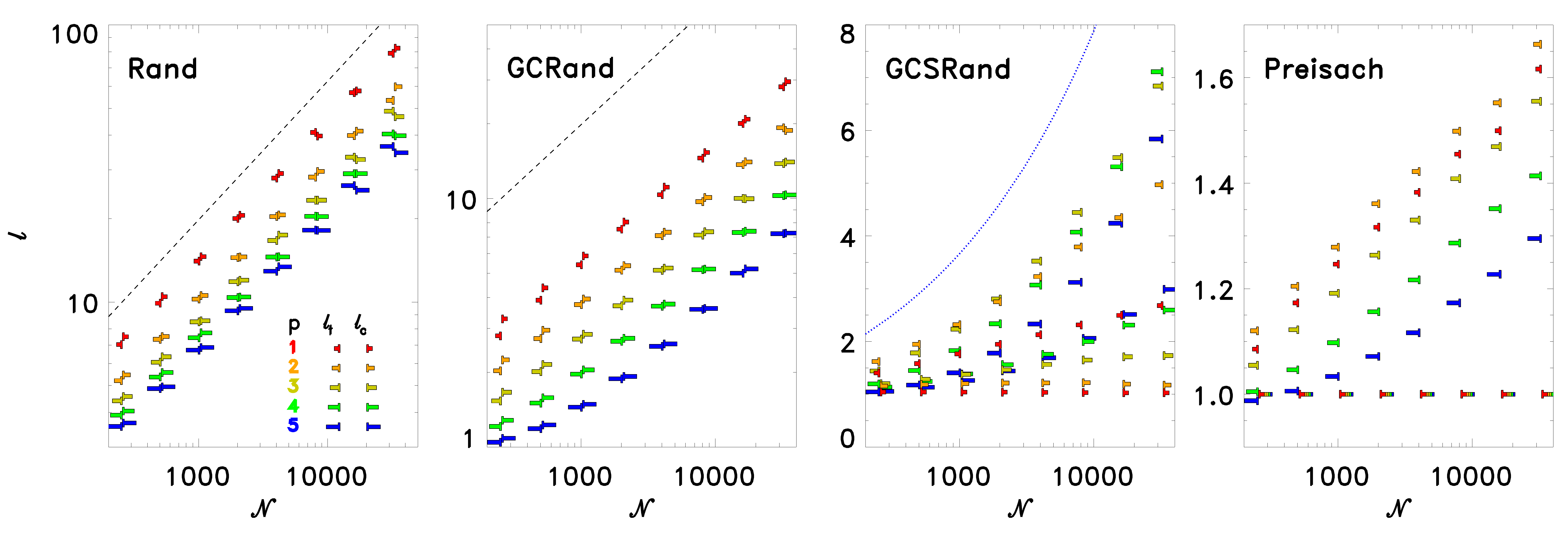} 
  \caption{Sample-averaged statistics of $\mathbf{R}_p = \Dn^p \Up^p$ for the random map classes {\tt Rand}, {\tt GCRand}, {\tt GCSRand} and {\tt Preisach} (from left to right column), introduced in Section \ref{sec:annealing}. 
 The average cycle lengths $\ell_c$ ($\vdash$) and transient lengths $\ell_t$ ($\dashv)$  are plotted against $\mathcal{N}$. 
  The color and size of the symbols correspond to the different values of $p$, as indicated in the left-most plot. The black dashed lines are the predictions for the simple random map, \eqref{eqn:ellc}, the blue dotted line in {\tt GCSRand} is a power law with exponent  $1/3$ and serves as a guide to the eye. } 
  \label{fig:num_data_ell}
\end{figure*}

    We begin with {\tt Rand} (first column). In this case $\mathbf{R}_p$ is the composition of the $p$th powers of two random maps. The component size $s_c$ (second row) does not change with $p$ and agrees very well with the random map prediction of \eqref{eqn:sc}, indicated by the black dashed line. The number of components $n_c$ on the other hand, does depend on $p$, decreasing with increasing $p$. Nevertheless, comparing with the prediction for the random map \eqref{eqn:sc}, the leading order behavior of $(\ln \mathcal{N})/2$ seems to be preserved and there appears to be an additive correction whose value  depends on $p$.  
    
    Comparing with the results for {\tt GCRand} in the second column, we see that the introduction of generation compatibility has modified the logarithmic dependence of the number of components $n_c$ on $\mathcal{N}$. We observe larger values of $n_c$, but as in the case of {\tt Rand}, $n_c$ remains small compared to the system sizes $\mathcal{N}$ with the powers $p$ considered. 
    
    For {\tt GCRand} the component size $s_c$ does also show a dependence on $p$. However, this dependence diminishes with increasing $\mathcal{N}$. The fact that the number of components scales as $\ln \mathcal{N}$ while $s_c$ scales as $\mathcal{N}$, suggests that the component size distribution is broad \footnote{Recall that the averages for $s_c$, $\ell_t$ and $\ell_c$ are averages weighted by the component size. Thus $s_c$ is the second moment of the component size distribution.}. Indeed, the typical functional graphs contain  
    few components whose size is comparable to the system size and many others that are much smaller. This is already apparent in Fig.~\ref{fig:random_graph}. In fact, for the simple random map Flajolet and Odlyzko \cite{FlajoletOdlyzko1990} predict that the expected size of the largest component scales with $\mathcal{N}$ as $s_{\rm max}/\mathcal{N} = 0.75782$. Similarly, for {\tt Rand} we find $s_{\rm max}/\mathcal{N} \approx 0.8$, with no discernible dependence on $p$ or $\mathcal{N}$, while for {\tt GCRand} and for the values considered $s_{\rm max}/\mathcal{N}$ approaches $0.2$ both for large $\mathcal{N}$ and large $p$. This is consistent with the $s_c$ behavior of {\tt GCRand}. This suggests that the introduction of generation compatibility in {\tt GCRand} has reduced the expected size of the largest component, but its size still scales linearly with $\mathcal{N}$. 
    
    The behavior of $n_c$ and $s_c$ for {\tt GCSRand} and {\tt Preisach} (third and fourth columns), while similar to each other, is markedly different from both {\tt Rand} and {\tt GCRand}. The number of components $n_c$ is much larger and proportional to $\mathcal{N}$ at large values of $p$ (note that $y$-axes for both maps are on a logarithmic scale). In both {\tt GCSRand} and {\tt Preisach}, we find that at constant $p$, $n_c$ appears to increase with $\mathcal{N}$ according to a power-law and the scaling exponent has a weak dependence on $p$. The corresponding component sizes $s_c$ change  in a reciprocal manner and this behavior is similar for {\tt GCSRand} and {\tt Preisach}. We also find that for both {\tt GCSRand} and {\tt Preisach} the ratios $s_{\rm max}/\mathcal{N}$ do not remain constant, but instead decreases with $\mathcal{N}$ for all $p$. We thus see that while for {\tt Rand} and {\tt GCRand} the functional graphs of $\mathbf{R}_p$ are dominated  by a few large components whose size remains comparable to $\mathcal{N}$, the restrictions leading to the maps {\tt GCSRand} and {\tt Preisach} change this picture drastically. The number of components for both are now comparable to the system size $\mathcal{N}$ and correspondingly, the component size distribution is less broad. 

    \subsection{Transient and cycle length} The numerical results for the transient and cycle lengths, $\ell_t$ and $\ell_c$ are shown in Fig.~\ref{fig:num_data_ell}.  Recall that for the simple random map $\ell_t = \ell_c$, given by \eqref{eqn:ellc}. For both {\tt Rand} and {\tt GCRand} this seems to be retained, the apparent deviations being of the same order as the statistical fluctuations. Likewise, the leading-order $\sqrt{\mathcal{N}}$-dependence survives, but with a $p$-dependent pre-factor. As one would expect, for fixed $\mathcal{N}$, $\ell_t$ and $\ell_c$ decrease monotonically with increasing $p$.

    \paragraph*{Cycle length} For {\tt GCSRand} and {\tt Preisach} the situation is again different. First of all,  $\ell_t \neq \ell_c$. In fact, for given $\mathcal{N}$ and $p$ we have $\ell_t > \ell_c$. For {\tt Preisach} moreover we find that $\ell_c = 1$ for all $p$, exactly. This is a direct consequence of the RPM property, as discussed at the end of Section \ref{sec:samples}. For {\tt GCSRand} and $p = 1$ we find that $\ell_c$ is slightly greater than one.   
    With $\mathcal{N}$ fixed, $\ell_c$ grows with increasing $p$. Within statistical precision this behavior appears to be monotonic. 
    
    \paragraph*{Transient length} We consider next the transient lengths $\ell_t$.  During the transient, the range of $\gamma$ covered need not be the same for every cycle, as noted in Section \ref{sec:samples}.  Thus the RPM property, which is based on a fixed range of $\gamma$, must be used with caution.  Remarkably, for both {\tt GCSRand} and {\tt Preisach}, at fixed $\mathcal{N}$ and increasing $p$, the behavior of $\ell_t$ is non-monotonic. For {\tt Preisach} $\ell_t$ increases for $p = 1, 2$ and thereafter starts to decrease, while for {\tt GCSRand} and depending on $\mathcal{N}$ the maximum is attained at $p = 2$ or $p =3$. For {\tt Preisach} the {\em maximum} transient length trivially satisfies $\ell_t \le L$, as is readily checked \footnote{In fact, one 
    can actually show that $\max \ell_t \leq  {\rm int}(L/2)$, where ${\rm int}(x)$ is the greatest integer less or equal to $x$, and the maximum is over all initial states and random realizations of the pair of maps.}. The observed {\em average} transient length is well less than that; $\ell_t < 2$ for the range of $\mathcal{N}$ and $p$ considered. Similarly, for {\tt GCSRand} we observe mean transient lengths less than  $7$ annealing cycles. Thus for both {\tt GCSRand} and {\tt Preisach}, $\ell_t \ll \mathcal{N}$, in contrast with {\tt Rand} and to a lesser extent {\tt GCRand}.     
    
    The non-monotonic behavior of $\ell_t$ with $p$ observed in the maps {\tt GCSRand} and {\tt Preisach} is worth commenting on. We consider first the decrease in $\ell_t$ with increasing $p$ (at fixed $\mathcal{N}$). Recall that in all the generation-compatible maps we considered, $\Up$ acting on a node $A$ reduces its $\Up$-generation $g$ by one, while $\Dn$ increases it by one. Thus the map $\mathbf{R}_p = \Dn^p \Up^p$ acting on a node $A$ of generation $g$, is effectively a map from the set of nodes of generation $g$ into itself. Clearly, given a node of generation $g$, the probability that it is revisited upon the next annealing cycle depends on the number of nodes $\mathcal{N}(g,L)$ in generation $g$. In general it will also  depend on the number of nodes in generations $g-1, g-2, \ldots g - p$, since $\mathbf{R}_p = \Dn^p \Up^p$ moves the node up and down $p$ generations (for simplicity we are assuming that $g + p < L$).

    It is not hard to see how the contraction factor can influence the length of a transient. Consider the extreme case where $\mathcal{N}(g + p,L) = 1$. Then starting with any node $A$ in generation $g$, $\Up^p A$ maps it to the single node of generation $g + p$. The subsequent map $\Dn^p$ brings it back to some node $B$ in generation $g$. Note however, that at this point a limit cycle of length one has been established: starting from $A$ and under repeated applications of $\mathbf{R}_p$, we have $A \to B \to B \ldots $. In going from {\tt GCRand} to {\tt GCSRand} to {\tt Preisach}, by imposing further restrictions, we effectively make the maps $\Up^p$ and $\Dn^p$ more contracting. This in turn tends to reduce transient and cycle length.

    \paragraph*{non-monotonicity} It is harder to understand the non-monotonicity in $p$ observed for {\tt GCSRand} and {\tt Preisach}. We believe that this arises out of an interplay of $p$ with $\mathcal{N}(g,L)$ as follows: Recall that for both types of maps the nodes can be interpreted as spin configurations. The spin configurations of  $\Up A$ and $\Dn A$ differ from those of $A$ in a single spin flip, those of $\Up^2 A$ and $\Dn^2 A$ in two flips {\em etc}. Thus for small values of $p$, the probability that $\mathbf{R}_p A = A$ will also depend on the number of available spins to flip, \ie the generation $g$ that $A$ belongs to.
    
    By our choice of 
    $\mathcal{N}(g,L)$ as a binomial factor,  \eqref{eqn:binom},  a large number of nodes are assigned to generations 
    $g \sim L/2$. To lowest order the image of nodes in the most populous generations determines the map statistics. Pick a node $A$ belonging to generation $L/2$ and consider its images under the map $\mathbf{R}_p = \Dn^p \Up^p$. For small values of 
    $p$, the number of nodes $\mathcal{N}(L/2+p,L)$ in generation $L/2 +p$ will be comparable to those in generation $L/2$ and thus the length of the transient will be dominated by the probability that spins flipped in the course of applying $\Up^p$ are undone by a subsequent application of $\Dn^p$. For small values of $p$ this probability is expected to decrease and the expected transient length therefore increases with $p$. 
    
    On the other hand, when $p$ becomes sufficiently large so that $\mathcal{N}(L/2+p,L)$ is much smaller than $\mathcal{N}(L/2,L)$, a focusing due to the large contraction factor of the map $\Up^p$ comes into play. The easiest 
    way to see this is to consider the equivalent map $\Up^p \Dn^p$ that maps the set of nodes in generation $L/2 + p$ into itself. Since $\mathcal{N}(L/2+p,L)$ decreases rapidly with increasing $p$ this will also reduce the expected transients length of this map, with the extreme case occurring when $\mathcal{N}(L/2+p,L) = 1$, as discussed before. 
    
    These types of ideas are very similar to those developed by Coppersmith in the context of Kauffman-networks  \cite{Sue2007} and the problem of determining whether a node is the image of some other node. We leave a more thorough analysis of our map classes for future work.

    \section{Discussion}\label{sec:discussion}

    Here we address the motivating question of the paper:  Can the convergence to periodic behavior observed in cyclic  annealing of bead packs be captured using generic discrete maps?  Apart from the mere existence of convergence, we also ask whether the convergence behavior of maps matches that of cyclic annealing \cite{Regev2015, Sastry2017}.  This convergence behavior includes the behavior of the transient length $\ell_t$ and the cycle length $\ell_c$ with the shear amplitude.  In bead pack annealing 
    these two lengths are typically comparable and they grow with increasing shear amplitude.  For systems of a few thousand particles $\ell_t$ and $\ell_c$ are of order $10$ cycles or less, hence $\ell_t$ and $\ell_c$ show no dramatic dependence on system size, which is consistent with similar 
    findings \cite{Sastry2017}. One also observes sub-synchronous motion when the shear amplitude approaches the ``yielding transition" at which convergence to cyclic behavior is lost.

    Here we investigated this question using arbitrary maps on an $\mathcal{N}$-object set.  The $\mathcal{N}$ objects represent all the locally stable configurations of a packing---exponential in the number of particles $N$. 
    The image of object $i$ under the map is the new configuration reached upon changing the shear amplitude until an instability occurs.  We distinguished two maps; one map $\Up$ dictating the transitions encountered on one direction of shear and the other ($\Dn$) dictating transitions under shear in the opposite direction.  We restricted these maps in various ways and observed the effect on their convergence properties.  In these studies the number of mapped states or configurations $\mathcal{N}$ was small relative to the physical systems we aim to describe.  Moreover, the size of the annealing cycles was kept small relative to those observed in physical systems.  The number of elementary transitions or glitches in an annealing cycle of our maps never exceeded ten, while the number observed in  \cite{Regev2015, Sastry2017} reached many dozens.  

    At the most qualitative level, these maps produce convergence to a periodic cycle of states that repeats every one or more annealing cycles in accord with observations.  The convergence length increases sublinearly with the number of states $\mathcal{N}$, in $\sqrt \mathcal{N}$ steps or fewer.  However, the convergence showed two major differences from the packings.  First, the  convergence was qualitatively unphysical.  It was either much too fast (a single cycle for {\tt Preisach}) or much too slow. Second, the convergence was typically faster for larger-amplitude cycles (larger $p$), in contrast to the packing simulations.  
    This is natural since each increase in $p$ dictates a reduction in the remaining active states by a contraction factor $z$, and thus larger-amplitude annealing cycles produce a larger net contraction factor for an annealing cycle; hence, faster convergence.

    The restricted maps that we used, {\tt GCRand}, {\tt GCSRand} improved the agreement with the packings modestly.  They reduced the transient lengths $\ell_t$.  
    Unlike the packings, the transient length still grew superlinearly in the number of degrees of freedom $N$ ($\goesas\log \mathcal{N}$), as indicated by the dashed lines in 
    Fig.~\ref{fig:num_data_ell}.  As for {\tt Preisach}, its transient length was of order unity for all of our runs.  When used with conventional annealing between two limiting $\gamma$'s, its intrinsic transient length is precisely unity, in view of its return-point memory property.  Thus the Preisach limit excludes the possibility of realistic annealing. 

    One promising feature of our map approach is the natural appearance of the sub-synchronous behavior seen in packing simulations.  Indeed, strongly sub-synchronous behavior covering many annealing cycles with $\ell_c >>1$ was the rule rather than the exception in our mappings.  We only observed the desired simple synchronous behavior ($\ell_c = 1$) when $p$ was large enough to approach saturation of the $\Up$ and $\Dn$ trees.  

    Much of the discrepancy between the map picture and the packing annealing stems from identifying $\mathcal{N}$ as the (exponentially large) number of configurations in the system.  Using this basis for choosing $\mathcal{N}$ clearly gives transient lengths that are enormously longer than is seen in the packings.  If the mapping picture is to have any resemblance to the packing behavior, the effective number of accessible configurations must be qualitatively less than the total number of configurations.  Since the observed convergence rates seem roughly independent of system size, the map picture can only be applicable if the effective $\mathcal{N}$ is also roughly independent of the size of the system.  

    The effective $\mathcal{N}$ of a map need not be comparable to the number of configurations.  Certainly if a map  happens to map small subsets of the configurations into themselves, then its effective $\mathcal{N}$ is the size of these subsets.  More generally, any map that tends to revisit states recently visited is more likely to repeat itself after a small number of iterations. The tendency of a map 
    to revisit recently visited states can be thought of as a form of ``locality". If the number of accessible states grows with the number of mappings performed so that the maps become increasingly non-local, one then expects effective $\mathcal{N}$ to grow as well.  Thus, expanding the annealing cycle (increasing the number of iterations $2p$ in a cycle) would tend to slow the convergence to a periodic behavior, as observed in the packings. The transition to non-convergence as the strain amplitude reaches $\gamma^*$ can be understood within the map picture as an increase of non-locality thereby causing an increase in the 
    effective $\mathcal{N}$ of the map.

    The maps we devised did restrict the possible set of states reachable under iteration.  For example in  {\tt GCRand}  $D^p U^p $ maps from a given generation $g$ to itself.  This restriction limits the accessible states to a fraction of $\mathcal{N}$, but not to a number independent of $\mathcal{N}$.  Thus it cannot give the fast convergence rates observed.  But other restrictions {\em can} provide much smaller effective $\mathcal{N}$.  If, for example, there exists a measure of distance between the states, then one can restrict the maps to be local with respect to this distance.  Such a restriction is not unphysical, since glitches are often observed to be localized disturbances in the sample.  

    \section{Conclusion} \label{sec:Conclusion}

    The picture introduced above aims to lay a common descriptive basis for explaining the remarkable convergence to periodic motion seen in many irreversibly annealed systems.  As we have seen, one can depict the annealing process via our abstract map language and compare the annealing behavior observed with that of realistic systems.  
    The maps are a useful way to capture the determinism of the annealing process. They clarify the inevitability of the convergence to a cyclic behavior whenever the iteration of the map leads to a repetition.  Understandably, the minimal maps we used do not achieve realistic convergence behavior.  This shows that the realistic convergence requires further physical restrictions, like the spatial locality mentioned above.  Exploring such restrictions is a promising path to gain insight about how synchronous and sub-synchronous behavior arise.  Conversely, one may use the packing simulations to gain information about the actual maps governing annealing for both geometric packing models and explicit discrete-state models\cite{Sethna93}.  We look forward to exploring these avenues in future work.

    \paragraph*{Acknowledgments} MM would like to thank J. Krug for useful discussions and for pointing out reference  \cite{DerridaFlyvbjerg1987}. The authors would like to thank Karin Dahmen, Nathan Keim, Sid Nagel, Joey Paulsen, Ido Regev, Srikanth Sastry, Ken Sekimoto, Mert Terzi and Lev Truskinovsky for many stimulating discussions. Nathan Keim's thoughtful suggestions on a previous draft led to major improvements in the present manuscript.  MM  was supported by the German Research Foundation (DFG) under DFG Projects No. 398962893 and the DFG Collaborative Research Center 1060 ``The Mathematics of Emergent Effects". This work was inspired in great measure by the winter 2018 program MEMFORM18 sponsored by the Kavli Institute for Theoretical Physics and funded in part by the US National Science Foundation under award PHY 17-48958.

\bibliography{cycles_maps}%
					
\end{document}